# Low-Impact Air-to-Ground Free-Space Optical Communication System Design and First Results

Alberto Carrasco-Casado
Ricardo Vergaz
José M. Sánchez-Pena

Universidad Carlos III de Madrid
EPS, Dpto. Tecnología Electrónica
Madrid, Spain
aacarras@ing.uc3m.es

Eva Otón
Morten A. Geday
José M. Otón

Universidad Politécnica de Madrid
ETSIT, Dpto. Tecnología Fotónica
Madrid, Spain
eva.oton@upm.es

*Abstract*—An air-to-ground free-space optical communication system has been designed and partially developed. The design covers both the communications between the airborne and the ground station, and the acquisition, tracking and pointing. A strong effort has been made in order to achieve the minimum payload power, size and weight, for which a MEMS modulating retroreflector has been chosen. In the ground station, a new technique for fine pointing, based on a liquid crystal device, is proposed and will be demonstrated, as well as other improvements with the aim of optimizing the ground station performance.

*Free-space optical communications; free-space lasercom; modulating retroreflector; retromodulator; acquisition, tracking and pointing, liquid-crystal-based beam steering.*

## I. INTRODUCTION

An air-to-ground free-space optical communication system has been designed and partially developed as an improvement to classic radio frequency (RF) links from Unmanned Aerial Vehicles (UAV) to ground stations. This project belongs to the SINTONIA program (acronym in Spanish for low-environmental-impact unmanned systems), led by BR&TE (Boeing Research and Technology Europe) with the purpose of boosting Spanish UAV technology. The work of GDAF-UC3M is under the coordination of INDRA, S.A. and INSA, S.A.

In the future, small UAVs will become an important part of national security. This kind of aircraft has proven to be tremendously flexible devices and can be used for many civil and military purposes. Telecommunication plays a more important role in the operation of UAVs than it does for manned aircraft since all the decision-making occurs on the ground (either before or during the flight). Currently, communications between UAVs and ground stations are based on RF systems and low-earth-orbiting satellite links [1]. Both are long range communications but also have low bandwidth (usually in the order of hundreds of kbps or less). The move to optical carrier frequencies means a qualitative leap because it provides a shift from MHz to hundreds of thousands of GHz,

This work has been supported by INSA (Ingeniería y Servicios Aeroespaciales S.A.), within the SINTONIA project (Sistemas No Tripulados Orientados al Nulo Impacto Ambiental), CENIT-E 2009 program, funded by CDTI (Centro para el Desarrollo Tecnológico Industrial) and partly by Comunidad Autónoma de Madrid: CAM (grant. nº. S2009/ESP-1781, FACTOTEM-2).

lowering the signal divergence by five orders of magnitude. Lower divergence allows higher reception power and signal-to-noise ratio, enabling faster communications with lower bit-error-rates [2]. It is also a more secure technique since the laser communication beam can not be intercepted without being noticed, as interception leads to signal fading. Another big advantage is that this technology enables lower power consumption and smaller and lighter terminals [3]. Since a first-class goal in this program is developing low-impact technology onboard UAVs, optical communications fit ideally to these requirements.

## II. REMOTE TERMINAL DESIGN

Given the strong effort needed to optimize the telecommunications payload, a modulating retroreflector (MRR) has been proposed as a communications remote terminal (Fig. 1).

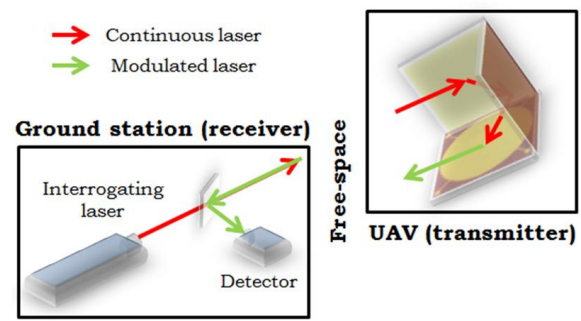

Figure 1. Modulating retroreflector principle

This device, which combines a retro-reflector and an optical modulator, is capable of returning light from a distant interrogating laser source without any additional pointing requirement onboard, while simultaneously modulating its intensity on the way back [4]. Such a device allows both the laser transmitter subsystem and the acquisition, tracking and pointing (ATP) subsystem to be fully eliminated on one end of the link, which results in a considerable reduction of power, size and weight onboard the UAV. The burden moves to the ground station but the ATP subsystem is eased since the MRR acts as a pointing reference by reflecting the income laser beam back to its source.

The viability of lasercom with minimum communication payload onboard UAVs (expected to weigh less than 1 kilogram and take up a few centimeters) is going to be demonstrated –for which MEMS-based (MicroElectroMechanical Systems) MRR will be used in the order of hundreds of kbps– as well as high-rate communications with UAVs –for which the ground laser transmitter will be OOK-modulated in the order of hundreds of Mbps.

## A. MEMS-based modulating retroreflector

In the past, experiments were made by the authors [6] testing liquid crystal technology as a transmissive modulator in an MRR scheme. Although it has great advantages such as its low consumption and light weight, the slow time response has proven to be a handicap difficult to overcome, only reaching speeds in the order of kHz.

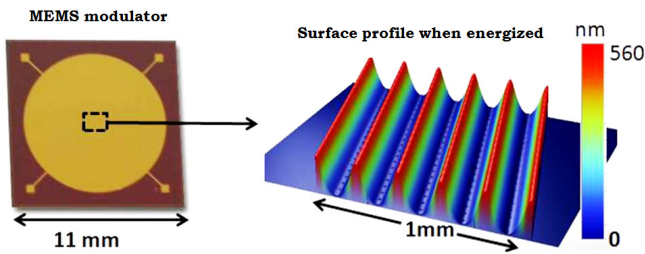

Figure 2. MEMS Optical Modulator and surface profile measurement (from [5])

The active device in the MRR designed is an MEMS modulator developed by Boston Micromachines Corporation (BMC). This modulator is a reflective diffraction grating with controllable groove depth (Fig. 2). It is one of the three mirrors that make up the retroreflector and is capable of modulating a continuous laser beam by switching between an unpowered flat-mirror state and an energized-diffractive state.

The modulator is fabricated on a conductive substrate that functions as one electrode of an array of elongated electrostatic actuators. The mirror surface acts as the other electrode, which is manufactured using MEMS technology and is suspended and electrically isolated from the substrate by an array of anchor supports. With the application of a voltage between the two electrodes, the actuators experience deflection corrugating the mirror surface. A trade-off exists between modulation contrast and dynamic response, so several devices will be tested throughout the MRR development, in order to optimize the communications performance, as well as other improvements such as different levels of vacuum environments during MRR fabrication so that the squeeze film damping effect can be minimized and speed maximized.

## III. GROUND STATION DESIGN

In the Fig. 3, a block diagram of the ground station can be seen. In the following sections, a more detailed description of the different subsystems will be made.

## A. Acquisition, tracking and pointing

The ATP subsystem is a key part of the ground station, since the remote terminal is airborne, with a wide range of acceleration and speed. It is based on a two-axis gimbal for tracking and coarse pointing, and a fast steering mirror (FSM) for fine pointing. The most widely used strategy in lasercom is a beacon-based ATP technique [7]. However, a beacon system is not suitable onboard the UAV, due to weight constraints. Nevertheless, a beacon-based system but with the laser in the ground station has been designed taking advantage of the retroreflector already installed onboard the UAV, with the advantages of a beacon system and with no extra payload. This strategy allows an optimal use of the MRR, acting as a communication terminal as well as a beacon reference for the ATP system.

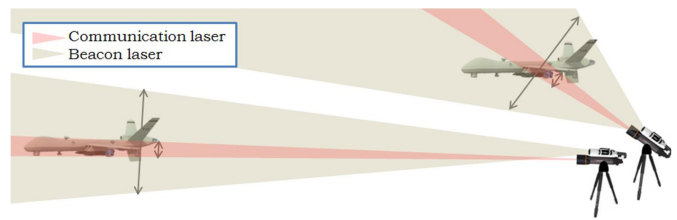

Figure 4. Beam divergence control to ensure a constant beam width regardless of the distance

The strategy designed for the ATP system is as follows: the pan-tilt gimbal continuously tracks the trajectory of the UAV, aided by its GPS position, and transmits an infrared beacon laser with a beam width wide enough to make up for the GPS error, so that the UAV is always illuminated by the beacon. This spot is estimated to be around 5 meters regardless of the distance. The gimbal is manufactured by FLIR Motion Control Systems and is a high-speed (up to 100º/s) system, capable of positioning a heavy load (<30 kg) with a resolution of 0.0064°. Based on the retroreflector principle, the reflection of the beacon can be constantly monitorized through an IR camera in order to determine the exact position of the UAV. As the distance between the two terminals is known through their GPS coordinates and its variability is high, a variable zoom lens will be needed in order to optimize the signal-to-noise ratio received by the IR camera, as well as a narrow optical filter to reject the wavelengths different from the beacon.

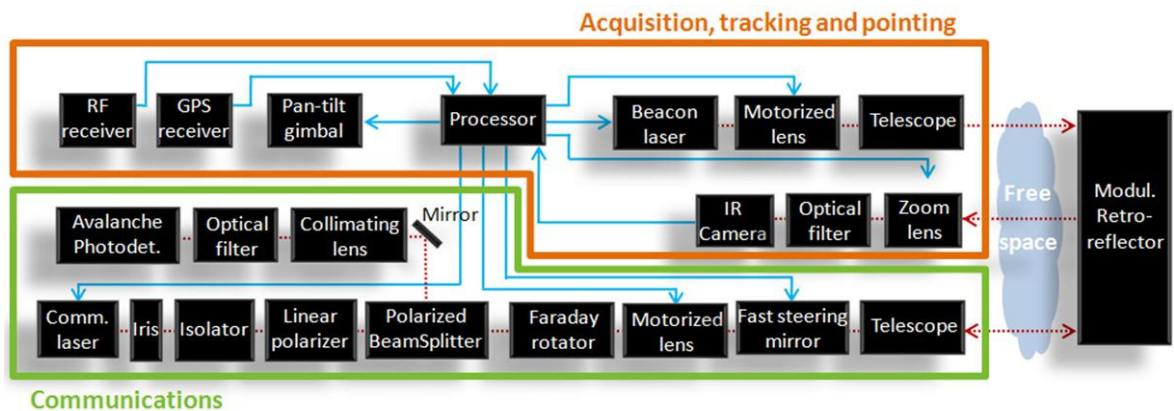

Figure 3. Optical ground station block diagram

The communication laser, with a much smaller (10-20 times) beam width than the beacon laser (Fig. 4), is transmitted through a telescope controlling its fine position with the FSM. Fine pointing is based on a biunivocal correspondence between the position of the beacon image in the camera focal plane and the movement of the FSM that is needed to illuminate the MRR with the communication laser. This strategy makes it possible to maintain a real-time fine pointing with the UAV in a continuous mode.

### B. Laser beam- divergence control

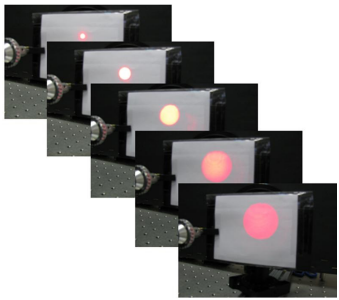

Figure 5. Laser beam-width control

Since the UAV-to-ground station distance may vary greatly, a beam-divergence control has been designed to produce the same beam width in the UAV regardless of the distance. This control is based on a motorized lens, which movements manage to defocus a laser that is collimated in the initial position, producing any desirable width beyond the difraction-limited one (Fig. 5).

The laser-beam width reaching the UAV needs to be designed carefully, as there is a trade-off between optical power and spatial coverage. The beacon laser has to be wide enough to make up for the GPS-position uncertainty in order to illuminate the retroreflector continuously. Regarding the communication laser, this width has to be small enough to achieve the needed signal-to-noise ratio, taking into account the way back from the retroreflector, but big enough to make up for the movements of the UAV that the FSM can not correct in time.

In the design of this beam width, it is also important to consider the FSM range of allowed angular laser movements before leaving the telescope and the minimum divergence achievable considering the largest link distance. Considering an initial maximum distance of 1000 meters, according to the diffraction limit of an aperture at 1550 nm, a minimum beam width at the telescope would be 2 cm if the goal is to achieve a coverage of ~10 cm reaching the remote terminal. It is important to note that in the distance range of an UAV, the beam width reaching the airborne is the result not only of the laser divergence, but also of the initial aperture, (which is the beam width at 0 meters), negligible if the distance is longer (Fig. 6). For distances longer than 1000 meters, the larger collimated beam width leaving the telescope makes up for the shorter range of FSM angular range needed to achieve the same resolution, avoiding using a larger telescope.

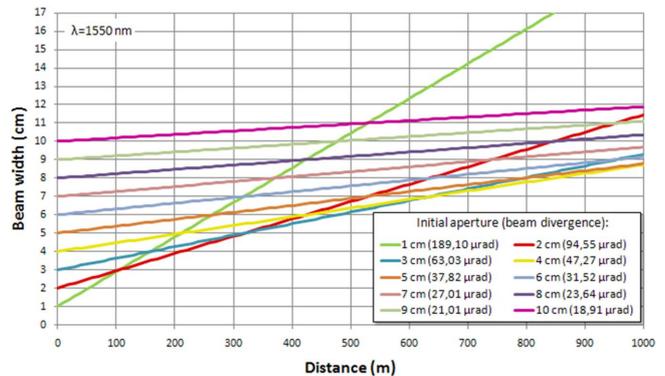

Figure 6. Beam width vs distance for a 1550 nm laser

### C. Polarization discrimination

An optical communication link based on the MRR scheme implies that the optical axis is the same for the transmitted and the received beam, assuming that the beam width is always bigger than the retroreflector aperture (otherwise the received beam would have an unknown offset from the interrogating laser optical axis, as it would depend on which of the three mirror faces the beam sees first). The simplest solution to enable the reception of the laser on its way back is to shift the ground receiver out of the optical axis, but an important part of the returning light is lost this way.

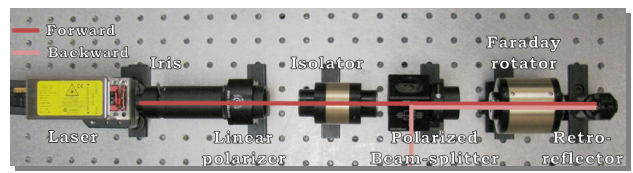

Figure 7. Polarization discrimination setup

The solution proposed in this design is creating two different optical paths within the same physical path. Since the wavelength is always the same, a spectral beamsplitter can not be used. The simplest way to receive in the same physical path is by means of a 50-50 beam splitter, by deflecting the laser on its way back to a different direction towards the photodetector. However, this method has an important drawback, as 50% of the optical power is lost in each pass. The technique proposed allows to minimize the power lost by transmitting in a linear polarization and receiving in the orthogonal polarization on the way back. This can be accomplished with a setup consisting of a Faraday rotator –since 45º linear polarization is rotated in each of the two passes– and a polarized beam splitter –which deflects the beam to a different direction when aligned with the orthogonal polarization (Fig. 7). A gain of ~7dB has been measured this way compared with the 50-50 beam splitter setup.

*D. Liquid-Crystal fine pointing experiment*

A novel device will be tested for fine pointing as an alternative to a FSM. This liquid crystal-based beam steerer is the only non-off-the-shelf component of the system, as it is manufactured by the coauthors of the paper, at the Universidad Politécnica de Madrid. It is based on a controllable orientation gradient of liquid crystal molecules. The device consists of two patterned Indium-Tin Oxide (ITO) coated glass plates assembled in a sandwich-like cell filled with a nematic liquid crystal (Fig. 8). The pattern is made up of high density (in the order of hundreds) transparent conductive ITO electrodes. The high resolution area is covered with a high resistivity coating (PEDOT), so interference problems are avoided. A voltage gradient on one of the substrates generates a graded switching profile across the cell, hence a graded refractive index. By changing the applied voltage, the liquid crystal orientation gradient changes and thus the steering angle is tunable.

The device offers a high accurate control of the laser beam angle in two dimensions with low voltages (~10 V). Since it is an analog device, the maximum angular resolution achievable is only limited by the electronics. It has the great advantage of being all-electro-optical instead of the electromechanical FSM, frequently used for fine pointing. The optical nature of the liquid-crystal cell behaviour makes it much less sensitive to external vibrations than an electromechanical one. The device will also allow the addition of tunable lens operation, so that the beam-divergence control would be integrated with the fine pointing, being the system a fully non-mechanical one. Such a device has a great potential in space applications and will be demonstrated within this project to test its capability within an actual free-space communication system.

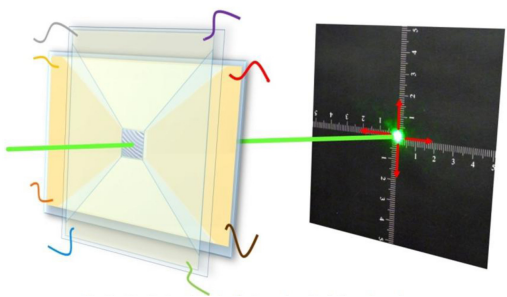

Figure 8.  2D laser beam steering

## IV. CURRENT STATE AND FUTURE WORK

A design of the lasercom system has been made and the ground station has been partially developed (Fig. 9). The full development of the ground station including the integration of all the subsystems will involve the main workload in the project and is to be done. At the moment, a laboratory test bed is being prepared including a trial for every part of the system, as well as a test of the final integrated system. An elevated rail layout has been built in the laboratory along which a robot will be travelling. The robot will carry the MRR, and its movements will be equivalent to the ones of a distant UAV. A miniature retroreflector will make it possible to simulate a long-distance experiment in the fine-pointing test. A series of field tests are also programmed, involving the use of a small radio-controlled airplane to begin with, and eventually an actual UAV for the final tests, the use of which is being negotiated.

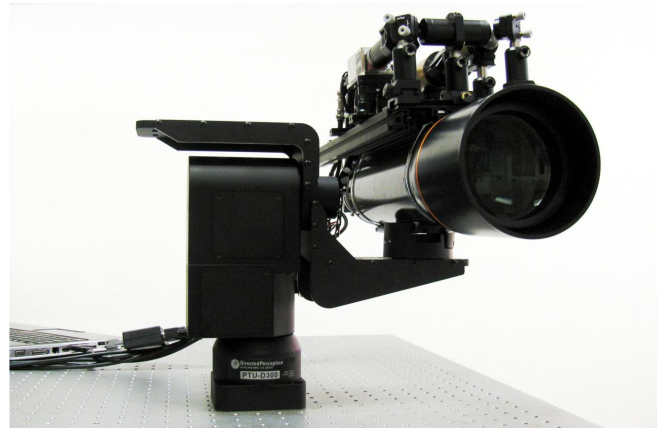

Figure 9.  Current state of the partially-developed ground station

Future system updates will include higher data rates using fiber optics technology at ground station, atmospheric turbulence mitigation with aperture averaging and adaptive optics, and active inertial stabilization of the ground station to make it suitable for mobile platforms.

ACKNOWLEDGMENT

We thank CENIT-E 2009 program, funded by CDTI (Centro para el Desarrollo Tecnológico Industrial, grant. CEN-20091001), and Comunidad Autónoma de Madrid: CAM (grant. nº S2009/ESP-1781, FACTOTEM-2).